\documentclass[11pt]{article}
\usepackage{amsfonts}
\usepackage{amsmath,amssymb}

\newcommand{\pa}{\partial}
\newcommand{\n}{\nonumber\\}

\newcommand{\fn}{\footnotesize}

\newcommand{\LLL}{\mathcal{L}}

\newcommand{\bec}{\begin{center}}
\newcommand{\eec}{\end{center}}

\newcommand{\bea}{\begin{array}}
\newcommand{\ear}{\end{array}}

\newcommand{\h}{\star_h}

\newcommand{\eps}{\vcx}

\newcommand{\bfr}{\begin{flushright}}

\newcommand{\efr}{\end{flushright}}
\newcommand{\noi}{\noindent}\newcommand{\Ra}{\rightarrow}
\newcommand{\me}{\frac{1}{2}}

\newcommand{\RR}{\mathbb{R}}

\newcommand{\ap}{\alpha}
\newcommand{\be}{\beta}

\newcommand{\la}{\Lambda}
\newcommand{\bege}{\begin{equation}}
\newcommand{\enge}{\end{equation}}
\newcommand{\w}{\wedge}
\newcommand{\g}{\gamma}

\newcommand{\si}{\sigma}

\newcommand{\beq}{\begin{eqnarray}}\newcommand{\benu}{\begin{enumerate}}\newcommand{\enu}{\end{enumerate}}
\newcommand{\eeq}{\end{eqnarray}}
\newcommand{\mt}{\mathcal}

\newcommand{\ee}{{\bf e}}
\newcommand{\jj}{{\bf j}}

\newcommand{\mr}{\mathring}

\newcommand{\ww}{{\bf w}}
\newcommand{\CC}{\mathbb{C}}

\newcommand{\ep}{\epsilon}

\newcommand{\vbn}{\blacktriangleleft}
\newcommand{\vvn}{\blacktriangleright}

\newcommand{\xx}{{\bf x}}

\newcommand{\bx}{\begin{pmatrix}}
\newcommand{\ex}{\end{pmatrix}}
\newcommand{\vcx}{\varepsilon}

\usepackage{indentfirst}

\begin{document}
\title{Spacetime Deformations and Electromagnetism in Material Media}
\author{{\bf R. da Rocha}\thanks{Instituto de F\'{\i}sica Gleb Wataghin (IFGW), Unicamp, Cidade Universit\'aria Prof. Zeferino Vaz, 
CP 6165, 13083-970, Campinas (SP), Brazil.
E-mail: {\tt roldao@ifi.unicamp.br}. Supported by CAPES.}\and {\bf Igor
Leite Freire}
 \thanks{Departamento de Matem\'atica Aplicada, IMECC, Unicamp, CP 6065, 13083-859, Campinas (SP), Brazil. E-mail:
{\tt igor@ime.unicamp.br}. Supported by CAPES.}}

\date{}\maketitle

\abstract${}^{}$\begin{center}
\begin{minipage}{12cm}$ \;\;\;\;\;${\footnotesize{
This paper is intended to investigate the relation between electrodynamics in anisotropic material media and its analogous 
formulation in an spacetime, with non-null Riemann curvature tensor. After discussing the electromagnetism via chiral differential 
forms, we point out the optical activity of a given material medium, closely related to 
topological spin, and the Faraday rotation, associated to topological torsion. Both quantities are defined in terms 
of the magnetic potential and the electric and magnetic fields and excitations. We revisit some properties of material media
and the associated Green dyadics. Some related features of ferrite are also investigated. It is well-known that the constitutive 
tensor is essentially equivalent to the Riemann curvature tensor. In order to investigate the propagation of 
electromagnetic waves in material media, we prove that it is analogous to consider the electromagnetic wave propagation in the vacuum, but this time
in a curved spacetime, which is obtained by a deformation of the Lorenztian metric of Minkowski spacetime. Spacetime deformations
 leave invariant the form of Maxwell equations.
Also, there exists a close relation between Maxwell equations in curved spacetime
and in an anisotropic material medium, indicating that electromagnetism and spacetime properties are deeply related.
 For instance, the equations of holomorphy in Minkowski spacetime are essentially Maxwell equations in vacuum.
Besides, the geometrical aspects of wave propagation can be described by an effective geometry which represents a modification of the Lorentzian metric 
of Minkowski spacetime, i.e.,
a kind of spacetime deformation.}}
\end{minipage}\end{center}
\medbreak
\medbreak\noi
\medbreak\noi  MSC classification: 15A03, 15A90, 34B27
\medbreak\noi Pacs numbers: 02.40.-k, 03.50.De, 11.15.Ex, 11.30.-j

\section*{Introduction}

The metric-free formulation of electromagnetism is an old concept 
\cite{ton,bal,punt,hehl1,janc2,kiehn,post,sal,scho},
  and it is
more natural, correct, precise and geometrically sensible if differential forms, intrinsically endowed with chirality \cite{schou,janc2}, are used
\cite{bay,post}. 
The excitations\footnote{We follow the nomenclature given in \cite{hehl1}.} {\bf D}$(x)$, {\bf H}($x$) (respectively the electric displacement and the magnetic field) 
 and the fields {\bf E}($x$), {\bf B}($x$) (respectively the electric field and the magnetic induction)  are naturally  described
in a spacetime destituted of a metric, which is introduced only when the constitutive relations are to be defined.
The metric-free (and affine-free) formulation of electrodynamics brings a geometric character and a clear physical interpretation.
But if we want to relate the Faraday excitation  2-form field $G(x)$ and the electromagnetic field strenght  2-form field $F(x)$, 
we need to consider a constitutive tensor $\chi$,
that gives the relation $G(x) = \chi F(x)$. Equivalently, spacetime must be endowed with a metric, since
the constitutive law depends on the spacetime metric \cite{punt} .

The constitutive tensor (CT) is more than a relation between $F(x)$ and $G(x)$, describing physical intrinsic properties of matter
 or spacetime. It is essentially the Hodge star operator \cite{jad} that, besides the well-known duality between $k$-forms and $(n-k)$-forms in a $n$-dimensional 
vector space (endowed with a metric), changes the
parity of differential forms, but with an additional information about the medium structure.
In this sense, CT describes the properties of spacetime (magnetic) permeability and (electric) permittivity, since a general medium  can be arbitrarily anisotropic.
Under this viewpoint, a CT immediately brings a light-cone conformal structure to spacetime \cite{jad}.
Formally, the link between CT and spacetime structure is expected, since CT and the Riemann curvature tensor have the same mathematical properties. 
A CT can reveal precious informations about spacetime,
  for example,
the CT scalar curvature  is identically null in
any medium possessing central symmetry \cite{post}.

In this paper the CT $\chi$ that describes any linear (in particular, crystalline) media,
 in the general case presenting optical activity, is expressed by a conformal transformation of the
vacuum CT $\chi_0$.
 The metric associated
to the medium is derived from the CT (up to a conformal factor) and vice-versa \cite{gross,hehl1,hehl2,hehl4}.
In this sense, the Lorentzian metric of Minkowski spacetime, associated with $\chi_0$, is deformed into a general metric of a riemannian
spacetime, related to $\chi$.

We prove that in order to completely describe the CT of any linear medium presenting natural optical activity,
 we only need the matrix $\gamma$, that describes the optical activity of such a medium, and $\chi_0$.
In the particular investigation of crystalline media,
we describe the constitutive tensor associated to the 32 crystal classes presenting natural optical activity
uniquely from  $\chi_0$, i.e., from the spacetime metric, since $\chi_0$ can be written as second order metric tensor combination \cite{hehl1,gross}.
This paper is organized as follows: in Sec. 1 we review the geometric description of fields and
 excitations in electromagnetic theory, well-established originally in the papers by
Maxwell \cite{max}, Sch\"onberg \cite{scho},  Hehl \cite{hehl1}, Kiehn \cite{kiehn}, Post \cite{post} and Jancewicz \cite{janc1,janc2},
among others. The Maxwell equations carry information about the nature of
the different fields and excitations, and the theory arises with an essential geometric character, if the fields and excitations
are correctly interpreted, as in \cite{janc1,janc2,hehl1,post}, as even and odd differential
form fields.  In Sec. 2 we revisit the main features of electromagnetism in anisotropic material media. The Green dyadic 
is obtained and we treat the particular case of a electrically anisotropic material media. Plane waves
are investigated, illustrating the present approach. Also, optically active media are treated. For instance, ferrite is explicitly studied.
In Sec. 3  Maxwell equations in curved spacetime
and in an anisotropic material medium are proved to be equivalent. It sheds some new light on the 
differential geometric aspects of electromagnetic wave propagation, that can be described by a geometry which represents a modification of the metric in Minkowski 
spacetime.
We prove that all the information contained in the constitutive tensor associated to any linear media (arbitrarily describing optical activity), 
is precisely given by the {\it vacuum}
constitutive tensor and by the matrix that describes the optical activity associated with the medium. Such matrix is essentially viewed as a perturbation 
of the original constitutive tensor that does not describe optically active media.  
Finally, in the Appendix, the main results concerning differential forms are reviewed.


\section{Electromagnetism in the exterior algebra}

Heretofore $\Omega^k(M)$ denotes the space of $k$-form fields 
defined on a manifold $M$. Given the map $ {\bf E}: M\rightarrow \RR^{3}$, the electric field ${\bf E}(x)$, $x\in M,$
 is an even 1-form field  ({\bf E}($x$) $\in \Omega^1(\RR^3)$),
since {\bf E}($x$) is a  linear map from the infinitesimal vector $d{\bf r}(x)$ to the infinitesimal scalar
potential $dV(x)$, given by $dV(x) = - {\bf E}(x)\cdot d{\bf r}(x)$. The physical dimension
of ${\bf E}(x)$ in the SI, [{\bf  E}($x$)] = Vm$^{-1}$, agrees with this interpretation. Analogously the
 magnetic induction ${\bf B}(x)$  is an
even 2-form field ({\bf B}($x$) $\in \Omega^2(\RR^3))$, since  {{\bf B}}($x$) is a linear map from the infinitesimal bivector $d{ \mt S}(x)$ to the
infinitesimal  scalar
 $d\phi(x)$. Explicitly we have  $d\phi(x) = - { \bf B}(x)\cdot d{\mt S(x)}$, where $\phi(x)$ is the magnetic flux. The physical dimension
${\bf B}(x)$ in the SI, [{\bf B}($x$)] = Wbm$^{-2}$ = T ( = Tesla), again agrees with such an intepretation \cite{max,janc2}.

From now on we call an even (odd) differential form field the one that doesn't (does) change sign under 
parity transformations\footnote{A parity transformation is defined in $\RR^n$ as the inversion of an odd number of basis vectors of $\RR^n$.} \cite{schou,janc2}.   
Even form fields are elements of $\Omega_+(M)$, hereon simply denoted by $\Omega(M)$,
while odd form fields are elements of $\Omega_-(M)$. Such forms are called \emph{chiral} differential forms.

\subsection{Homogeneous Maxwell equations and potentials}
The differential operator $d:\Omega^p_\pm(M)\Ra\Omega^{p+1}_\pm(M)$
  does not change the differential forms parity. From now on we denote
$\pa_\ap = \pa/\pa\ap$. We also adopt natural units, in particular the speed of light $c=1$, in what follows.

The first homogeneous Maxwell equation is an expression
relating even 2-form fields:
\bege\label{hom1}
d{\bf E}(x) + \pa_t {\bf B}(x) = 0.
\enge
\noi The absence of magnetic monopoles can be described by the equation
\bege\label{dbo}
d{\bf B}(x) = 0.
\enge\noi Eqs.(\ref{hom1}) and (\ref{dbo}) are the homogeneous Maxwell equations. From eq.(\ref{dbo}),
 using the Poincar\'e lemma\footnote{The Poincar\'e lemma asserts that if an open set $\mho\subset\RR^n$ is {\it star-shaped}, 
every closed form is exact in $\mho$. ($\omega\in\Omega^k (M)$ is closed if $d\omega = 0$, and exact if there exists a form $\eta\in\Omega^{k-1}$
such that $\omega = d\eta$. An open set $\mho \subset \RR^n$ is {star-shaped} with respect to the origin if,
 for all $\xx\in\mho$, the line from the origin to $\xx$ is in $\mho$.)}, there exists an even 1-form field
 {\bf A}($x$) (the magnetic potential) satisfying the relation
\bege
{\bf B}(x) = d{\bf A}(x).
\enge\noi Substituting in eq.(\ref{hom1}), one obtains the expression $d{\bf E}(x) + \pa_t d{\bf A}(x) = 0$, or
 $d\,({\bf E}(x) + \pa_t{\bf A}(x)) = 0$. Using the Poincar\'e lemma, there exists a scalar field potential $\Phi(x)\in\Omega^0(\RR^3)$
 such that ${\bf E}(x) + \pa_t{\bf A}(x) = - d\Phi(x)$, implying that
\bege
{\bf E}(x) = -\pa_t {\bf A}(x)  - d\Phi(x).
\enge

\subsection{Non-homogeneous Maxwell equations}

The electric current density $\jj(x)$ is an element of
$\Omega^2_-(\RR^3)$, an odd 2-form field, which changes sign under parity transformations. It is clear that
$d\jj\in\Omega^3_-(\RR^3)$, and then the continuity equation (local form of electric charge conservation) can be written as
\bege\label{cce}
d\jj(x) + \pa_t \rho(x) = 0,
\enge
\noi where $\rho(x)$, the electric charge density, is an odd  3-form field.
Obviously $d\rho(x) = 0$, and the Poincar\'e lemma again asserts that there exists ${\bf D}(x)\in\Omega^2_-(\RR^3)$ such that
\bege\label{pol}\rho(x) = d{\bf D}(x).\enge \noi The
1-form field {\bf D}($x$) is called {electric displacement}.
The unity of {\bf D}($x$) in the SI is [{\bf D}($x$)] = Cm$^{-2}$ (C = Coulomb).
Substituting in eq.(\ref{cce}), we have $d\jj(x) + \pa_t (d{\bf D}(x)) = 0$, and $d\,(\jj(x) + \pa_t{\bf D}(x)) = 0$.
There exists an odd 1-form field ${\bf H}(x)\in \Omega^1_-(\RR^3)$ such that
\bege\label{pom}
\jj(x) + \pa_t{\bf D}(x) = d{\bf H}(x).\enge
\noi It describes the Amp\`ere-Oersted law. The SI unit of {\bf H}($x$) is [{\bf H}($x$)] = Am$^{-1}$ (A = Amp\`ere).
The odd form fields  ${\bf D}(x)$ and ${\bf H}(x)$ are potentials with sources $\rho(x)$ and $\jj(x)$, respectively.

The Poyinting vector {\bf S}($x$) describes the electromagnetic strenght  energy flux density.
It is possible to write
\bege
{\bf S}(x) = {\bf E}(x)\w{\bf H}^*(x).
\enge\noi From the algebraic viewpoint, the product above is the unique possibility, since quantities representing flux densities are
 described  by odd differential 3-form fields \cite{hehl2,janc2,schou,misner,postl}, elements of $\Omega^3_-(\RR^3)$.
The electric ($\ww_e(x)$) and  magnetic ($\ww_d(x)$) energy densities are elements of $\Omega^3_-(\RR^3)$, expressed by
\bege
\ww_e(x) = \frac{1}{2}{\bf E}(x)\w{\bf D}(x) = \frac{1}{2}{\bf D}(x)\w{\bf E}(x),\quad \ww_m = \frac{1}{2}{\bf B}(x)\w{\bf H}(x) = \frac{1}{2}{\bf H}(x)\w{\bf B}(x).
\enge
\noi The electromagnetic field energy density is written as \cite{janc2}
\bege
\ww = \ww_e + \ww_m = \frac{1}{2}({\bf E}(x)\w{\bf D}(x) + {\bf B}(x)\w{\bf H}(x)).
\enge

\subsection{The Hodge star operator}
It is well-known \cite{bt} that the vector spaces  $\Omega^k(\RR^3)$ and
 $\Omega^{3-k}(\RR^3)$ have the same dimension, since 
$ {\rm dim} \;\Omega^k(\RR^3) = {3 \choose k} = {3 \choose 3-k} =  {\rm dim}\; \Omega^{3-k}(\RR^3).$ 
 The same result is valid to any $n$-dimensional space (see Appendix). 
 Meanwhile, it does not exist any canonical  isomorphism between $\Omega^k(\RR^3)$ and $\Omega^{3-k}(\RR^3).$
 The  isomorphism given by the {\it Hodge star operator}  $\star : \la^k(V) \rightarrow \la^{n-k}(V)$ always satisfies  $\star \star= \pm id$.
For more details, see the Appendix. The contraction is a generalization of the interior product, and it can be written in terms of the Hodge star 
operator and the exterior product, as
\begin{equation}
\psi\lrcorner \phi=\star((\star \phi)\wedge \psi),
\end{equation}\noi where $\psi,\phi\in\Omega(\RR^3)$.
For more details see, e.g., \cite{bt,bur}.

\subsection{The Poynting theorem}
If we take the $\CC$-conjugation of eqs.(\ref{hom1}) and (\ref{pom}) and respectively multiply by ${\bf E}^\ast(x)$ and
 ${\bf H}^{\ast}(x)$ we obtain
\beq\label{464} {\bf E}(x)\w d{\bf H}^{\ast}(x)-{\bf H}^{\ast}(x)\w d{\bf
E}(x) &=& \frac{1}{c}\Big{[}{\bf E}(x)  \w ({\pa}_{t}\star \vcx
{\bf E} ^{\ast}(x))+\star \mu^{-1}{\bf B}(x) \w ({\pa}_t{\bf
B}^{\ast}(x)) \Big{]}\n &=&\star \Big{[} {\bf E}(x) \lrcorner
\;({\pa}_t\vcx {\bf E}^{\ast}(x)) +(\mu^{-1}{\bf B})(x)\lrcorner
\;({\pa}_t{\bf B}^{\ast}(x)) \Big{]}. \eeq
\noi  Eq.(\ref{464}) can be written in a coordinate system as
\begin{equation}
\nonumber \frac{1}{2}{\partial_t}\Big{(} {\bf E}
(x)\lrcorner(\vcx {\bf E}(x)) + {\bf B}(x)\lrcorner
(\mu^{-1}{\bf B}(x)) \Big{)}dx \w dy \w dz.
\end{equation}
\noi and  from the expression
\begin{equation}
-d{\bf S}(x)={\bf H}^{\ast}(x)\w d{\bf E}(x) -{\bf E}(x) \w d{\bf H}^{\ast}(x),
\end{equation}
\noi it follows that
\begin{equation}
 d{\bf S}(x)=\partial_t \ww(x),
\end{equation}\noi 
the so-called \textit{Poynting theorem} \cite{heav,bay,born}.

\subsection{Electromagnetic Intensity and Excitation}
 The electromagnetic field strenght  $F(x)\in
\Omega^2(\RR^{1,3})$ is an even 2-form in $\RR^{1,3}$, also  called the \emph{Faraday 2-form field} \cite{misner}.
If an arbitrary, but fixed, time vector is chosen in $\RR^{1,3}$, we can split spacetime in space plus time.
Then it is possible to use ${\bf E}(x)$ and ${\bf B}(x)$ to describe $F(x)$ as
\bege
 F(x) = {\bf B}(x) + {\bf E}(x)\w dt.
\enge\noi

The electromagnetic excitation  $G(x)\in \Omega^2_-(\RR^{1,3})$ can also be considered as an odd 2-form field  given by \cite{janc2,hehl2}
\bege
 {G}(x) = {\bf D}(x) - {\bf H}(x)\w dt.
\enge
\noi Eqs.(\ref{hom1},\ref{dbo}) can be summarized as
\bege\label{cuz}
dF(x) = 0,\enge
\noi and eqs.(\ref{pol},\ref{pom}) are synthetically  written as
\bege
d{G}(x) = {J}(x),\enge
\noi when the odd 3-form current density field  $ {J}(x) = \rho(x) - \jj(x)\w dt $ is defined \cite{janc2}.
If we admit primarily eq.(\ref{cuz}), the electric and magnetic fields are only defined after a spacetime splitting.

\subsection{Vacuum constitutive relations}

Hereon it is assumed implicitly that the Hodge star operator \emph{changes} the parity of the
differential forms\footnote{This Hodge star operator is, \emph{de facto}, the composition of  the Hodge star operator
 with a pseudoscalar \cite{rota,bf1,bf2}. This new operator is then able to lead odd (even) form fields to even (odd) ones.
 (By abuse of notation we also denote this new operator 
by $\star$. }.

Constitutive relations are written as
 \bege G(x) = \star F(x). \enge\noi This relation can be expressed in the vacuum,  after
a spacetime splitting, as: \bege {\bf D}(x) = \vcx_0\star{\bf
E}(x),\qquad\qquad {\bf B}(x) = \mu_0\star{\bf H}(x), \enge \noi where
$\vcx_0$ denotes the vacuum electric permittivity and $\mu_0$ denotes the
vacuum magnetic permeability.
 From eq.(\ref{cuz}) it is possible to find ${A}(x)\in \Omega^1(\RR^{1,3})$ such that
\bege\label{uz}
F(x) = d{{A}(x)}.\enge\noi The even 1-form field $A(x)$ denotes the well-known electromagnetic potential.
In components, eq.(\ref{uz}) is written as \bege F_{\mu\nu} = \pa_\mu{{A}}_\nu - \pa_\nu{{A}}_\mu.\enge
Eq.(\ref{uz}) is invariant under the maps $A(x)\mapsto A(x) + \omega(x)$, $\omega(x)\in \Omega^1(\RR^{1,3})$ such that $d\omega(x) = 0$.
In particular eq.(\ref{uz}) is invariant  when ${A}(x) \mapsto {A}(x) + d\lambda(x), \;\; \lambda(x)\in \Omega^1(\RR^{1,3}).$
The existence of form fields that are closed, but {\it not} exact, gives rise to the physical monopole and solitons in fluids, concerning
paramount and striking applications such as  
superconductivity, topological defects and turbulent non-equilibrium thermodynamics of fluids, exhaustively investigated 
by Kiehn \cite{kiehn}. 

One \cite{kiehn} defines the  odd 3-form field topological spin $S(x) = A(x)\w G(x)\in \Omega^3_-(\RR^{1,3})$ and the even 3-form
field topological torsion  $T(x) = A(x)\w F(x)\in \Omega^3(\RR^{1,3})$. It can be shown that optical activity is closely related to 
topological spin, while  Faraday rotation is associated to topological torsion \cite{kiehn}.

 Under a spacetime splitting it can be seen that
\beq\label{tt}
T(x) = A(x)\w F(x) &=& ({\bf A}(x) - \phi(x) dt)\w ({\bf B}(x) + {\bf E}(x)\w dt)\nonumber\\
  &=& {\bf A}(x)\w {\bf B}(x) + ({\bf A}(x)\w {\bf E}(x) - \phi(x) {\bf B}(x)) dt\eeq\noi and
\beq
S(x) = A(x)\w G(x) &=& ({\bf A}(x) - \phi(x) dt)\w ({\bf D}(x) - {\bf H}(x)\w dt)\nonumber\\
  &=& {\bf A}(x)\w {\bf D}(x) + ({\bf A}(x)\w {\bf H}(x) - \phi(x) {\bf D}(x)) dt\eeq\noi

Kiehn \cite{kiehn} shows that $T(x)$ is related to the helicity, while $S(x)$ is  associated to chirality
 of the electromagnetic fields.
The 3-form field energy-momentum, is defined if an arbitrary direction $\ee_i$ is chosen:
\bege
 U_i(x) = \me [F(x)\w (\ee_i\lrcorner G(x)) - G(x)\w (\ee_i\lrcorner F(x))].
\enge\noi The 3-form field energy-momentum is invariant under
pseudodual maps $F(x)\mapsto \varphi(x) G(x)$ and $G(x)\mapsto -
F(x)/\varphi(x)$, where $\varphi(x)$ is an arbitrary scalar field non-null in all points of $\RR^{1,3}$.

\section{Revisiting electromagnetism in non-homogeneous media}

In the last decade, a lot of manuscripts have been concerning electrodynamics in material 
media via differential forms. For instance, see
 \cite{war1,war2,war3,fre,janc1}.

For permeability and permittivity tensors such that the product
 $\eps^{t}\mu^{-t}$ is diagonalizable, the expression for the Green diadic \cite{war1,war2,war3} is given by 
\begin{equation}\label{po}
g=\frac{\det \mu}{4 \pi \tilde{r}}\begin{pmatrix}
  \exp({i m_{1}\tilde{r}}) & 0 & 0 \\
 0 &  \exp({i m_{2}\tilde{r}}) &0  \\
 0 & 0 &  \exp({i m_{3}\tilde{r}})
\end{pmatrix},
\end{equation}
where $\tilde{r}=\sqrt{\det \mu}\Big{(}\frac{x^{2}}{\mu_{1}}
+\frac{y^{2}}{\mu_{2}}+\frac{z^{2}}{\mu_{3}}
\Big{)^{\frac{1}{2}}}$, and $m_{1}$, $m_{2}$, $m_{3}$ denote eigenvalues 
associated with the matrix $\eps^{t}\mu^{-t}$, such that
$Re \text{ }m_{i}>0$, i=1,2,3.  $\mu_{1}$, $\mu_{2}$, $\mu_{3}$
denote eigenvalues associated with $\mu$.
\par
The diadic given by eq.(\ref{po})  can be immediately written as the tensor product \cite{war2,war3}:
\begin{equation}
g=\frac{\det \mu}{4 \pi \tilde{r}}\Big{(}e^{i m_{1}\tilde{r}}
dx^{1}\otimes dx^{2}+e^{i m_{2}\tilde{r}}dy^{1} \otimes
dy^{2}+e^{i m_{3}\tilde{r}}dz^{1} \otimes dz^{2} \Big{)}.
\end{equation}

\subsection{Material media}

A medium is completely isotropic if the electric permittivity $\eps$ and the magnetic permeability $\mu$ can be written, respectively,
as $\mu = $ diag($\mu_1, \mu_1, \mu_1$) and $\eps =$ diag($\eps_1, \eps_1, \eps_1$).
If the elements of $\mu$ and $\eps$ in an anisotropic medium are hermitian,  such medium is called {eletrically
or  magnetically girotropic}.
For instance, a plasma with static magnetic field in the $z$ axis
\begin{equation}\label{ma1}
\eps=\begin{pmatrix}
  \eps_{1} & -i \eps_{p}  &0  \\
  i \eps_{p} & \eps_{1} & 0 \\
  0 &0  & \eps_{z}
\end{pmatrix}
\end{equation}\noi is eletrically girotropic.  
A magnetic girotropic medium
is given by
\begin{equation}
\mu=\begin{pmatrix}
  \mu_{1} & -i \mu_{2}  & 0 \\
  i\mu_{2} & \mu_{1} & 0 \\
  0 & 0 & \mu_{z}
\end{pmatrix}.
\end{equation}
\subsection{Electrically anisotropic media}

Suppose that, in a given material medium, $\eps = {\rm diag} (\eps_{1},\eps_{2},\eps_{3})$ and $\mu = I$, where 
$I$ denotes the identity matrix.
It is immediate that $\tilde{r}=\mu_{1}r$ and $\det \mu=\mu_{1}^{3}$, from where it can be seen that
$\eps\mu^{-t}=\mu_{1}^{-1}$ diag($\eps_{1}, \eps_{2}, \eps_{3}$). Since the Green form $g$ is given by \cite{war2}
\bege\label{gre}
{g} = \left[(\det \mu)^{-1} k^t \mu k I - \omega^2 \eps \mu^{-t}\right]^{-1},
\enge\noi then the diagonal components of ${g}$ are given by 
\beq
{g}_{jj}(k)&=&\Big{[}\frac{k^{2}}{\mu_{1}^{2}}-\omega^{2}\frac{\eps_{j}}{\mu_{1}}\Big{]}^{-1}\n
&=&\frac{\mu_{1}^{2}}{k^{2}-\omega^{2}\eps_{j}\mu_{1}}.
\eeq
A solution of eq.(\ref{gre}) is given by  
\begin{equation}
{g}=  \frac{\mu_{1}^{2}}{4\pi r}\begin{pmatrix}
  \exp({i\omega \sqrt{\eps_{1}\mu_{1}}r}) &0  &0  \\
    0& \exp({i\omega \sqrt{\eps_{2}\mu_{1}}r})& 0 \\
    0 & 0& \exp({i\omega \sqrt{\eps_{3}\mu_{1}}r})
\end{pmatrix}
\end{equation}
and then the Green diadic is given by
\beq\label{tryu}
g(\vec{r}_{1},\vec{r}_{2})&=&\frac{\mu_{1}^{2}}{4 \pi r}\Big{(}\exp (i \omega
\sqrt{\eps_{1}\mu_{1}}r)\, dx_{1}\otimes dx_{2}+ \exp (i \omega
\sqrt{\eps_{2}\mu_{1}}r)\,dy_{1}\otimes dy_{2}\n
&& +\exp(i \omega
\sqrt{\eps_{3}\mu_{1}}r)\,dz_{1}\otimes dz_{2}\Big{)}
\eeq\noi
where $r=\|\vec{r}_{1}-\vec{r}_{2}\|$. This equation is originally obtained by Warnick \cite{war2}.

In the particular case of an uniaxial medium, with $\eps_{1}=\eps_{2}=\eps$ and $\eps_{3}=\eps_{z}$, eq.(\ref{tryu})
is led to
\beq g(\vec{r}_{1},\vec{r}_{2})&=&\frac{\mu_{1}^{2}}{4 \pi r}\Big{(}\exp (i \omega
\sqrt{\eps\mu_{1}}r)\, dx_{1}\otimes dx_{2}+ \exp (i \omega
\sqrt{\eps\mu_{1}}r)\,dy_{1}\otimes dy_{2}\n
&& +\exp(i \omega
\sqrt{\eps_z\mu_{1}}r)\,dz_{1}\otimes dz_{2}\Big{)}
\eeq\noi
In any material isotropic medium we have $\eps_{z}=\eps$, and  the vacuum is obtained when
$\eps\mapsto \eps_{0}$ and $\mu\mapsto\mu_{0}$. In this case, \bege\label{516}
g=\frac{\mu_{0}^{2}}{4\pi r} I
\enge
\subsection{Plane waves}
Heretofore we denote ${\bf E} = {\bf E}(x), {\bf B} = {\bf B}(x)$, \ldots, in order to simplify the notation to be used.
Suppose that a plane wave propagates in the ${s}$ direction
and let the electric field be expressed by
\begin{equation}\label{dept}
{\bf E} ={\bf E} _{0}\exp({i(k\vec{s}\cdot\vec{r}-\omega t)}).
\end{equation}
Denoting $ n={k}/{\omega} = 1/v$, eqs.(\ref{hom1}) and (\ref{pom}) gives
\begin{equation}\label{rf}
{\bf D}=-{n} {\bf H} \w s, \qquad
{\bf B}={n}{\bf E} \w s.
\end{equation}
Using the constitutive relation ${\bf B}=\h {\bf H}$,  eqs.(\ref{rf}) give
\beq\label{3}
{\bf D}&=&\frac{n^{2}}{\mu} \star(s\w {\bf E} )\w s\n
&=&\frac{n^{2}}{\mu}\star\Big{[} {\bf E} -s(s\cdot {\bf E} )\Big{]}.
\eeq\noi By abuse of notation, here $\mu$ denotes the magnetic permeability, a real number, instead of the tensor $\mu$. 
Define the component ${\bf E} _{\bot}$ of ${\bf E}$ such that ${\bf E} _\bot\cdot s = 0$. Then, since ${\bf E} _\bot$ is in the plaquette
defined by ${\bf E} \w s$, eq.(\ref{3}) is written as
\begin{equation}\label{plj}
{\bf D}=\frac{n^{2}}{\mu}{\bf E} _{\bot}.
\end{equation}
\subsection{Fresnel equations}
Now let $\eps_{1}$, $\eps_{2}$ and $\eps_{3}$ be the eigenvalues of the matrix representation of
$\vcx$. From eq.(\ref{3}) we have
\begin{equation}
{ E} _{i}=\frac{n^{2}s_{i}}{n^{2}-\mu \eps_{i}}{\bf E} \cdot s,
\end{equation}\noi from where we obtain
\beq\frac{s^{2}_{1}}{n^{2}-\mu \eps_{1}}{\bf E}\cdot s+\frac{s^{2}_{2}}{n^{2}-\mu \eps_{2}}{\bf E}\cdot
s+\frac{s^{2}_{3}}{n^{2}-\mu \eps_{3}}{\bf E}\cdot
s=\frac{1}{n^{2}}{\bf E}\cdot s \eeq

\begin{equation}\label{525}
\sum_{i=1}^{3}\frac{s_{i}^{2}}{n^{2}-\mu
\eps_{i}}=\frac{1}{n^{2}}.
\end{equation}
Since $s\cdot s = 1$, then $\sum_{i=1}^{3}s_{i}^{2}=1,$
and from eq.(\ref{525}) it follows that
\begin{equation}\label{526}
\sum_{i=1}^{3}\frac{s_{i}^{2}}{\frac{1}{n^{2}}-\frac{1}{\mu
\eps_{i}}}=0.
\end{equation}
\noi If we define the so-called principal propagation velocity $v_{i}:=({\mu \eps_{i}}^{-1/2}$, 0, 0), eq.(\ref{526}) is lead to
\begin{equation}\label{528}
\frac{s_{1}^{2}}{v^{2}-v^{2}_{1}}+\frac{s_{2}^{2}}{v^{2}-v^{2}_{2}}+\frac{s_{3}^{2}}{v^{2}-v^{2}_{3}}=0.
\end{equation}
 Eqs.(\ref{525}), (\ref{526}) and (\ref{528}) are called {Fresnel wave equations} \cite{born}.

\subsection{Ferrite}

Ferrite is a material medium defined by
\begin{equation}\label{532}
\mu=\mu_{0}\begin{pmatrix}
  \alpha & -i \beta  & 0 \\
  i\beta & \alpha & 0 \\
  0 &0  & \gamma
\end{pmatrix}=:\mu_{0}\mu_{r}
\end{equation}
From now on we consider $\beta<\alpha$, and it is easy to see that $\det
\mu=\mu_{0}^{3}(\alpha^{2}-\beta^{2})\gamma$ and that the eigenvalues of
$\mu_{r}$ are $\alpha+\beta$, $\alpha-\beta$ and $\gamma$. From the expression
\begin{equation}
\nonumber
\vec{\tilde{r}}:=(\alpha^{2}-\beta^{2})^{\frac{1}{2}}\gamma^{\frac{1}{2}}\mu_{0}\Big{(}\frac{x}{\sqrt{\alpha+\beta}},
\frac{y}{\sqrt{\alpha-\beta}}, \frac{z}{\sqrt{\gamma}} \Big{)},
\end{equation}\noi it follows that
\begin{equation}
\tilde{r}:= \|\vec{\tilde{r}}\| = \mu_{0}(\alpha^{2}-\beta^{2})^{\frac{1}{2}}\gamma^{\frac{1}{2}}\Big{(}
\frac{x^{2}}{\alpha+\beta}+\frac{y^{2}}{\alpha-\beta}+\frac{z^{2}}{\gamma}
\Big{)}^{1/2}.
\end{equation}
The square roots of the eigenvalues of 
 $\omega^{2}\eps \mu^{-t}$ are given by 
\begin{equation}
m_{1}=\omega \sqrt{\frac{\eps}{(\alpha+\beta)\mu_{0}}},\qquad
m_{2}=\omega \sqrt{\frac{\eps}{(\alpha-\beta)\mu_{0}}},\qquad 
m_{3}=\omega \sqrt{\frac{\eps}{\gamma\mu_{0}}},
\end{equation}
where $\eps$ is the diagonal element of the ferrite permittivity tensor. 
$g$ is given by
\begin{equation}\label{5}
g=\frac{\mu_{0}^{2}(\alpha^{2}-\beta^{2})^{\frac{1}{2}}\gamma^{\frac{1}{2}}}{4
\pi r^{'}}\;{\rm diag} (e^{i\omega \sqrt{\eps\mu_{0}(\alpha-\beta)\gamma}r\prime}, e^{i\omega \sqrt{\eps\mu_{0}(\alpha+\beta)\gamma}r\prime}, 
  e^{i\omega \sqrt{\eps\mu_{0}(\alpha^{2}-\beta^{2})}r\prime}),
\end{equation}
where
$\vec{\tilde{r}}=\sqrt{(\alpha^{2}-\beta^{2})\gamma}\mu_{0}\vec{r}^{\;\prime}$
and $\vec{r}^{\;\prime}=\Big{(}\frac{x}{\sqrt{\alpha +
\beta}},\frac{y}{\sqrt{\alpha - \beta}}, \frac{z}{\sqrt{\gamma}}
\Big{)}$. Eq.(\ref{5}) is equivalent to the expression
\begin{equation}
g(\vec{r}_{1},\vec{r}_{2})=g_{0}\Big{(}e^{i \omega
\sqrt{(\alpha-\beta)\gamma\eps \mu_{0}}r^\prime}dx^{1}\otimes
dx^{2}+e^{i \omega \sqrt{(\alpha+\beta)\gamma\eps
\mu_{0}}r^\prime}dy^{1}\otimes dy^{2}+e^{i \omega
\sqrt{(\alpha^{2}-\beta^{2})\eps \mu_{0}}r^\prime}dz^{1}\otimes
dz^{2} \Big{)}
\end{equation}
where
$g_{0}={\mu_{0}^{2}(\alpha^{2}-\beta^{2})^{\frac{1}{2}}\gamma^{\frac{1}{2}}}/{4
\pi r^\prime}$,
$r^{\prime}=\|\vec{r_{1}}^{\;\prime} -\vec{r_{2}}^{\;\prime}\|$ and
 $\vec{r}_{i}^{\;\prime}$ are analogously defined as
 $\vec{r}^{\;\prime}$. When  $\beta=0$ and $\alpha=\gamma$ in eq.(\ref{5}) it follows that
\begin{equation}\label{6}
g=\frac{\mu_{0}^2\alpha^{2}}{4 \pi r}
  \exp({i\omega \sqrt{\eps\mu_{0}\alpha} r}) I,
\end{equation}
where  $r_{i}=\sqrt{x_{i}^{2}+y_{i}^{2}+z_{i}^{2}}$.
Denoting $\mu=\alpha \mu_{0}$ eq.(\ref{6}) can be written as
\begin{equation}
g=\frac{\mu^{2}}{4 \pi r}  \exp({i\omega \sqrt{\eps\mu} r}) I,
\end{equation}
which is the well-known expression for an isotropic medium.

\subsection{Faraday rotations}

From eq.(\ref{pom}) and eq.(\ref{dept}), it follows that
\begin{equation}
\star d{\bf H} = -i\omega \eps {\bf E}.
\end{equation}
Taking the differential of the last equation we obtain
\beq\label{548}
d \star d {\bf H} &=& \omega^{2}\eps \h {\bf H}\n
&=&\omega^{2}\eps \mu {\bf H}.
\eeq
\noi Now, if we solve eqs.(\ref{rf}) for ${\bf B}$, we obtain
\bege\label{gv}
{\bf B} = \frac{n^2}{\eps}\star_h {\bf H}_\bot,
\enge\noi where ${\bf H}_\bot = {\bf H} - (s\cdot {\bf H}) s.$
In components, the field given by eq.(\ref{gv}) is written as 
\bege\label{541}
B_i = \frac{{n^2} \mu_i}{n^2 -  c^2\eps \mu_i} (s \cdot H)s_i.  
\enge
From eq.(\ref{541}) the component of ${\bf H}$ in the $z$-direction is zero, and if we make the assumption
that ${\bf H}=(H_{1}dx+H_{2}dy)e^{i(kz-\omega t)}$, eq.(\ref{548}) gives
\begin{equation}\label{551}
\omega^{2}\eps \mu_{0}\begin{pmatrix}
  \alpha & -i\beta \\
  i\beta & \alpha
\end{pmatrix}=k^{2}\begin{pmatrix}
  H_{1} \\
  H_{2}
\end{pmatrix}
\end{equation}
which solution is given by
\begin{equation}\label{552}
k^{2}_{+}=\omega^{2}\eps \mu_{0}(\alpha + \beta),\qquad
k^{2}_{-}=\omega^{2}\eps \mu_{0}(\alpha - \beta),
\end{equation}
describing two (left- and right-handed) circularly polarized plane waves.
Now, substituting eqs.(\ref{552}) in eq.(\ref{551}) it follows that
$H_{1}=\pm iA$, if  $H_{2}=A$, $A\in \CC$.
The general solution of the system is
\begin{equation}
{\bf H}=\Big{[} -ic_{1}A e^{ik_{+}z}+ic_{2}A
e^{ik_{-}z}\Big{]}e^{-i\omega t}dx+\Big{[} -c_{1}A
e^{ik_{-}z}+c_{2}A e^{ik_{-}z}\Big{]}e^{-i\omega t}dy.
\end{equation}
\noi Choosing $c_{1}=-c_{2}=\frac{i}{2}$, we obtain
\begin{equation}
H_{1}(z)=\frac{A}{2}e^{ik_{+}z}+\frac{A}{2}e^{ik_{-}z},\qquad
H_{2}(z)=i\frac{A}{2}e^{ik_{+}z}-i\frac{A}{2}e^{ik_{-}z},
\end{equation}\noi which can be written as
\beq
H_{1}(z)&=&A\; \cos\Big{(}\frac{k_{+}-k_{-}}{2}z
\Big{)}\exp(-i({k_{+}+k_{-}})z/2),\n
H_{2}(z)&=&A\; \sin\Big{(}\frac{k_{+}-k_{-}}{2}z
\Big{)}\exp(-i({k_{+}+k_{-}})z/2).
\eeq
\noi Let  $\theta\in \RR $ such that
\begin{equation}
\tan \theta=\frac{H_{2}(z)}{H_{1}(z)}=\tan \Big{(}
\frac{k_{+}-k_{-}}{2}z\Big{)}.
\end{equation}
It is immediate that
\begin{equation}
\theta_{k}=\left(\frac{k_{+}-k_{-}}{2}z\right) + 2k\pi,\quad  k\;\;\;\text{is an integer}.
\end{equation}
Restricting $\theta\in[0,2\pi)$ it is clear that  the phase difference between the left- and right-handed
components is 2$\theta_0$, where
\begin{equation}\label{563}
\theta_{0}=\frac{1}{2}z\omega \sqrt{\eps
\mu_{0}\alpha}\left(\sqrt{1+\frac{\beta}{\alpha}}-\sqrt{1-\frac{\beta}{\alpha}}
\right).
\end{equation}

Consider $k_\pm$ in eq.(\ref{552}) given by a second-order approximation, i.e.,
\begin{equation}
k_{\pm}=\omega \eps \mu_{0}\left(1\pm
\frac{\alpha}{2\beta}+\frac{1}{8}\frac{\alpha^{2}}{\beta^{2}} + {\mt O}\left(\frac{\ap}{\be}\right)^3\right).
\end{equation}
Substituting in eq.(\ref{563}) we have
\begin{equation}
\theta_{0}=\frac{\beta}{2\alpha}z\omega \sqrt{\alpha \mu_{0}\eps}.
\end{equation}
\noi It shows the well-known result asserting that ferrite is indeed a non-reciprocal medium.

 \section{Material media viewed as spacetime deformations in vacuum}

In the formalism in, e.g. \cite{hehl2,postl} that describes the electromagnetism in linear media, the dual Hodge operator action is equivalent to
the constitutive $\chi$ tensor action on 2-form fields:
\bege\label{clo}
\star\Delta = \chi\Delta, \qquad \Delta \in \Omega^2_\pm(\RR^{1,3}).\enge
If cartesian coordinates are introduced, eq.(\ref{clo}) is equivalent to
\bege
\star\Delta = \frac{1}{4}\ep_{\mu\nu\si\kappa}\chi^{\si\kappa\tau\rho}\Delta_{\tau\rho} dx^\mu\w dx^\nu.
\enge

\subsection{The constitutive tensor}
In this subsection we present and discuss the main results in, e.g., \cite{post,postl}, concerning 
the relation between the constitutive and the Riemann curvature tensors. 
In linear media, the 2-form electromagnetic intensity $F(x)\in\Omega^2(\RR^{1,3})$ is related
to the  electromagnetic excitation $G\in\Omega^2_-(\RR^{1,3})$ by the equation
\bege
F(x) = \chi G(x).
\enge\noi Using cartesian coordinates, $F(x)$ and $G(x)$ are expressed as
\bege
G(x) = \me G_{\mu\nu}(x) dx^\mu\w dx^\nu, \qquad F(x) = \me F_{\mu\nu}(x) dx^\mu\w dx^\nu
\enge
and  $F_{\mu\nu}(x)$ and $G_{\mu\nu}(x)$ are related by
\bege\label{0}
G_{\mu\nu}(x) = \frac{1}{4}\epsilon_{\mu\nu\ap\be}\chi^{\ap\be\si\lambda}F_{\si\lambda}(x),
\enge\noi where  $\epsilon$ is the  Levi-Civita tensor. The symmetry
\bege\label{1}
\chi^{\lambda\nu\si\kappa} = - \chi^{\lambda\nu\kappa\si}, \qquad \chi^{\lambda\nu\si\kappa} = - \chi^{\nu\lambda\si\kappa}.
\enge \noi arises, since $F_{\mu\nu}(x)$ and $G_{\mu\nu}(x)$ are antisymmetric. Besides, the lagrangian density $\LLL(x) = G(x)\w F(x)$
is written as \cite{post,postl}
\bege
\LLL(x) = \frac{1}{4} \chi^{\lambda\nu\si\kappa} F_{\lambda\nu}(x) F_{\si\kappa}(x).
\enge\noi From the relation
\bege
G^{\lambda\nu}(x) = 2 \frac{\pa\LLL(x)}{\pa F_{\lambda\nu}(x)} = \me \chi^{\lambda\nu\si\kappa} F_{\si\kappa}(x).
\enge\noi there exists the relation
\beq\label{2}
\chi^{\lambda\nu\si\kappa} &=& 2 \frac{\pa^2\LLL(x)}{\pa F_{\lambda\nu}(x) \pa F_{\si\kappa}(x)} = 2 \frac{\pa^2\LLL(x)}{\pa F_{\si\kappa}(x)
 \pa F_{\lambda\nu}(x)}\nonumber\\
&=& \chi^{\si\kappa\lambda\nu}\eeq\noi For more details, see \cite{postl}. 
 The number of independent coordinates is 21, (using eqs.(\ref{1},\ref{2})), which comes from the analogy to the Riemann curvature
 tensor. Only the antisymmetric combinations are non-trivial. An order two antisymmetric tensor has exactly six components and then
there would exist
36 componentes. Expressing $\chi$ as a $6\times 6$ matrix  ($\chi\in {\rm Hom}(\RR^6,\RR^6)$) and using the bivectorial notation,
 (where the indices $I, J,\ldots =
01,02,03,23,31,12$ are defined) we can see from eq.(\ref{2}) that the matrix $\chi^{IJ}$ is symmetric ($\chi^{IJ} = \chi^{JI}$), and
there exists (6$\times$7)/2 independent componentes in $\chi$.
Using physical arguments, one can show that in uniform media we have the relation \cite{post,postl}
\bege
\chi^{[\lambda\nu\si\kappa]} = 0.
\enge\noi In vacuum $\chi$ can be written as
\bege
\chi^{\lambda\nu\si\kappa} = Y_0 \sqrt{g} (g^{\lambda\si} g^{\nu\kappa} - g^{\lambda\kappa} g^{\nu\sigma}),
\enge\noi where $g$ is the determinant of $g^{\mu\nu}$ and $Y_0$ is the vacuum admittance.

From the constitutive tensor $\chi$ Post defines two invariants:
\bege
\chi_1 = \chi^{\;\;\;\;\lambda\nu}_{\lambda\nu},
\enge \noi which is called scalar curvature of the medium described by $\chi$, and
\bege
\chi_2 = \epsilon_{\lambda\nu\si\kappa}\chi^{\lambda\nu\rho\tau}\chi^{\si\kappa\mu\ap}\ep_{\rho\tau\mu\ap}.
\enge\noi Post [Po72] proves that $\chi_2$ is non-zero for any medium and $\chi_1 = \chi^{[\lambda\nu\si\kappa]}$ is identically null in
any medium possessing central symmetry. The constitutive tensor is explicitly represented by:
\medbreak
\begin{center}{\footnotesize{
$\chi$ = $\begin{pmatrix}-\vcx&\g\\
                                        \g^\dagger&\mu^{-1}\end{pmatrix}$ = \begin{tabular}{|r|rrr|rrr|}\hline
$\chi^{\lambda\nu\si\kappa}$ & 01 & 02 & 03 & 23 & 31 & 12\\
                       &$-E_1$&$-E_2$&$-E_3$&$B_1$&$B_2$&$B_3$\\\hline
   01 $D_1$             &$-\vcx_{11}$&$-\vcx_{12}$&$-\vcx_{13}$&$\g_{11}$&$\g_{12}$&$\g_{13}$\\
 02 $D_2$             &$-\vcx_{21}^*$&$-\vcx_{22}$&$-\vcx_{23}$&$\g_{21}$&$\g_{22}$&$\g_{23}$\\
 03 $D_3$             &$-\vcx_{31}^*$&$-\vcx_{32}^*$&$-\vcx_{33}$&$\g_{31}$&$\g_{32}$&$\g_{33}$\\\hline
23 $H_1$             &$\g_{11}^*$&$\g_{21}^*$&$\g_{31}^*$&$\zeta_{11}$&$\zeta_{12}$&$\zeta_{13}$\\
31 $H_2$             &$\g_{12}^*$&$\g_{22}^*$&$\g_{32}^*$&$\zeta_{21}^*$&$\zeta_{22}$&$\zeta_{23}$\\
12 $H_3$             &$\g_{13}^*$&$\g_{23}^*$&$\g_{33}^*$&$\zeta_{31}^*$&$\zeta_{32}^*$&$\zeta_{33}$\\

\hline\end{tabular}   } }
\end{center}
\noi The matrix $\mu_{lk}$ is the magnetic permeability matrix,  $\vcx_{lk}$ is the electric permittivity
matrix and $\g_{lk}$ is a matrix that describes the electric and magnetic polarization effects.
One can prove that in media possesing central symmetry, the matrix $\g_{lk}$
is null \cite{post,postl}. In isotropic media the relations
\bege
\g_{lk}\equiv 0, \qquad \vcx_{lk} = \vcx_0\delta_{lk}, \qquad \zeta_{lk} = \mu_0^{-1} \delta_{lk},
\enge\noi are satisfied. In this case, $\chi_1 = 0$ and $\chi_2 = -12\ep_0/\mu_0$ \cite{post}.

We shall study the light propagation in crystalline media presenting optical activity, which are characterized by  32 classes \cite{voi,post}.
 Each class is represented by a symmetry represented in the table:
\medbreak
\begin{center}{\footnotesize{
\begin{tabular}{||r|r||r|r||r|r||r|r||}\hline\hline
1 & $C$ & 9 &$C, z_2, x_2$&17&$C, z_4$&25&$z_6$\\\hline
2 & $-$ & 10 &$z_3, x_2$& 18&$z_4$&26& $z_3, x_2, E_z$\\\hline
3 & $C, z_2$& 11 & $z_3, E_x$&19& $S_z, x_2$, &27& $z_3, E_z$\\\hline
4 & $E_z$&12&$C, z_3$&20&$S_z$&28&$C, x_4, y_4$\\\hline
5 & $z_2$&13&$z_3$&21&$C, z_6, x_2$&29&$x_4, y_4$\\\hline
6 & $C, z_2, x_2$&14&$C, z_4, x_2$&22&$z_6, x_2$&30&$S_x, S_y$\\\hline
7 & $z_2, x_2$&15&$z_4, x_2$&23&$z_6, E_x$&31&$C, x_2, y_2, S$\\\hline
8 & $z_2, E_x$&16&$z_4, E_x$&24&$C, z_6$&32&$x_2, y_2, S$\\\hline
\hline\end{tabular}}}
\end{center}
\medbreak
\noi $C$ denotes central symmetry, $S$ is the cyclic permutation  of the indices, $E_x$ is a reflection with respect to the  $yz$ plane
 (analogous definition for $E_y$ and $E_z$), $S_x$ is a rotation using the $x$ axis, followed by a reflection
related to the  $yz$ plane (And analogous definitions for $S_y$ and $S_z$).

All crystal classes described in the above table present natural optical activity, and the corresponding respective matrices $\g_{kl}$, 
composing the tensor $\chi$, are described below (the number before the matrices indicates the class number above described):
$${{
 2\;\begin{pmatrix}
\g_{11}&\g_{12}&\g_{13}\\
\g_{21}&\g_{22}&\g_{23}\\
\g_{31}&\g_{32}&\g_{33}
\end{pmatrix};\quad
 4\; \begin{pmatrix}
0&0&\g_{13}\\
0&0&\g_{23}\\
\g_{31}&\g_{32}&0
\end{pmatrix};\quad
5\;\begin{pmatrix}
\g_{11}&\g_{12}&0\\
\g_{21}&\g_{22}&0\\
0&0&\g_{33}
\end{pmatrix};}}
$$$${{7\;\begin{pmatrix}
\g_{11}&0&0\\
0&\g_{22}&0\\
0&0&\g_{33}
\end{pmatrix};\quad
8\;\begin{pmatrix}
0&\g_{12}&0\\
\g_{21}&0&0\\
0&0&0
\end{pmatrix};\quad
10, 15, 22\;\begin{pmatrix}
\g_{11}&0&0\\
0&\g_{11}&0\\
0&0&\g_{33}
\end{pmatrix},}}
$$$$
{{11,16,23\;\begin{pmatrix}
0&\g_{12}&0\\
-\g_{12}&0&0\\
0&0&0
\end{pmatrix};\quad
13,18,25\;\begin{pmatrix}
\g_{11}&\g_{12}&0\\
-\g_{12}&\g_{11}&0\\
0&0&\g_{33}
\end{pmatrix};\quad
19\;\begin{pmatrix}
\g_{11}&0&0\\
0&-\g_{11}&0\\
0&0&0
\end{pmatrix};}}
$$$$
{{20\begin{pmatrix}
\g_{11}&\g_{12}&0\\
\g_{12}&-\g_{11}&0\\
0&0&0
\end{pmatrix};\quad
29,32\;\begin{pmatrix}
\g_{11}&0&0\\
0&\g_{11}&0\\
0&0&\g_{11}
\end{pmatrix}}}
$$
The matrices corresponding to the classes  29,32 describe the chiral vacuum \cite{kiehn}.

\section{Arbitrary constitutive tensors from the vacuum CT}
From now on we adopt the notation $F = F(x), G = G(x),\ldots$, omitting the argument $x$.

\medbreak
$\vvn$ {\bf Spectral Theorem}: There always exists a conformal transformation that diagonalizes the constitutive tensor $\chi$ $\vbn$
\medbreak  \noi Considering the splitting $\RR^{1,3} \simeq \RR^3\times \RR$, we can write
\bege
G = \chi F \Longleftrightarrow \binom{{\bf D}}{{\bf H}} = {{\begin{pmatrix}\label{p1}
-\vcx&\g\\\g^\dagger&\mu^{-1}
\end{pmatrix}_{6\times 6} }}\binom{-{\bf E} }{{\bf B}}
\enge\noi where
$ \vcx = \vcx_0\,I$ and  $\mu^{-1} = \mu_0^{-1}\,I$. By the theorem above, there exists a matrix $\Gamma$ composed by the eigenvectors of
 $\chi$ such that $\Gamma^{-1}\chi\Gamma = \Lambda$ is a diagonal matrix.

\subsection{The chiral vacuum}

In order to illustrate the general approach, 
we firstly consider the chiral vacuum, described by the matrix
{{\bege\label{m1}\begin{pmatrix}
-\vcx&\g_{\circlearrowleft}\\\g^\dagger_{\circlearrowleft}&\mu^{-1}
\end{pmatrix}_{6\times 6}\enge}} where $\g_{\circlearrowleft} = \g_{11}\,I$. The matrix (\ref{m1}) has eigenvalues $\si_1$, $\si_2$ and eigenvectors
 $\{(0,0,\si_1,0,0,1),\\ (0,\si_1,0,0,1,0), (\si_1,0,0,1,0,0), (0,0,\si_2,0,0,1), (0, \si_2,0,0,1,0), (\si_2,0,0,1,0,0)\},$ where
 \bege
\si_{1,2} = \mu_0^{-1} - \vcx_0 \pm \frac{\sqrt{(\mu_0^{-1} + \vcx_0)^2 - 4\g_{11}^2}}{2\g_{11}}.
\enge\noi   Then
\bege\Gamma^{-1}\chi\Gamma = \Lambda = {{\begin{pmatrix}-\Sigma_1&0\\0&\Sigma_2 \end{pmatrix}_{6\times 6}}}
                 \enge\noi where $\Sigma_1 = -\sigma_1\,I$ and $\Sigma_2 = \sigma_2\,I$.
Denoting  $\mr{F}  =\Gamma^{-1} F$ and $\mr{G} = \Gamma^{-1} G$, we obtain
\bege\binom{{\bf D}}{{\bf H}} = \chi \binom{-{\bf E} }{{\bf B}} \Rightarrow \Gamma\binom{\mr{{\bf D}}}{\mr{{\bf H}}} = \chi \Gamma\binom{-\mr{{\bf E} }}{\mr{{\bf B}}} \Rightarrow
\binom{\mr{{\bf D}}}{\mr{{\bf H}}} = \Gamma^{-1}\chi \Gamma\binom{-\mr{{\bf E} }}{\mr{{\bf B}}} =  {{\begin{pmatrix}-\Sigma_1&0\\0&\Sigma_2 \end{pmatrix}}}
\binom{-\mr{{\bf E} }}{\mr{{\bf B}}}.
\enge\noi Define the odd form fields \d{\it {\bf D}} and \d{\it {\bf H}} as
\bege
\d{${\bf D}$} = \Sigma_1^{-1}\mr{{\bf D}}, \qquad \d{${\bf H}$} = \Sigma_2^{-1}\mr{{\bf H}}.
\enge\noi It follows that
\bege
\binom{\d{\it {\bf D}}}{\d{\it {\bf H}}} = {{\begin{pmatrix}\Sigma_1^{-1}&0\\0&\Sigma_2\end{pmatrix}}} \binom{-\mr{{\bf D}}}{\mr{{\bf H}}} =
\begin{pmatrix}-1&0\\0&1\end{pmatrix} \binom{-\mr{{\bf E} }}{\mr{{\bf B}}},
\enge\noi and we prove that
\bege
\d{$G$} = \chi_0 {\mr{F}},
\enge\noi where $\chi_0$ vacuum constitutive tensor\footnote{modulo dilation of the axis $\ee_4, \ee_5$ and $\ee_6$ by $\mu_0$ and
contraction of $\ee_1, \ee_2$ and $\ee_3$ by $\vcx_0$.}. We obtain for the chiral vacuum, after doing the inverse maps, the constitutive
relation
\bege
G = \chi^{\circlearrowleft} F
\enge \noi where 
\bege
{\chi^{\circlearrowleft} = \Gamma\Lambda\chi_0\Gamma^{-1}}
\enge\noi Then the constitutive tensor $\chi^{\circlearrowleft}$, 
related to the chiral vacuum, is completely described  by the matrix $\g$  and the vacuum constitutive tensor.
We only used conformal maps in $\RR^{1,3}$, which are elements of the group. This kind of structure in electromagnetism was discovered by Bateman \cite{bate},
who was the first to observe that the Maxwell equations are invariant under the conformal group \cite{kiehn,whi}.

\subsection{Arbitrary linear media: crystalline media, optical activity, magnetic and dieletric Faraday effects}

The method is fundamentally analogous to the chiral vacuum case. Consider an arbitrary linear media described by the matrix
\bege\label{mm1}\chi = {{\begin{pmatrix}
-\vcx&\g\\\g^\dagger&\mu^{-1}
\end{pmatrix}_{6\times 6},}}\enge where
\bege
\g = {{\begin{pmatrix}
\g_{11}&\g_{12}&\g_{13}\\
\g_{21}&\g_{22}&\g_{23}\\
\g_{31}&\g_{32}&\g_{33}
\end{pmatrix}.}}
\enge\noi The matrix (\ref{mm1}) has eigenvalues $\si_A$ ($A = 1,2,\ldots, 6$). Then
\bege\Gamma^{-1}\chi\Gamma = \Lambda = {{\begin{pmatrix}-\Sigma_1&0\\0&\Sigma_2 \end{pmatrix}}}_{6\times 6}
                 \enge\noi where $\Sigma_1 = - {\rm diag}(\sigma_1, \si_2, \si_3)$ and $\Sigma_2 = -{\rm diag}(\sigma_4,\si_5, \si_6)$.
 Denoting $\mr{F}  =\Gamma^{-1} F$ and $\mr{G} = \Gamma^{-1} G$, we obtain
\bege\binom{{\bf D}}{{\bf H}} = \chi \binom{-{\bf E} }{{\bf B}} \Rightarrow \Gamma\binom{\mr{{\bf D}}}{\mr{{\bf H}}} = \chi \Gamma\binom{-\mr{{\bf E} }}{\mr{{\bf B}}} \Rightarrow
\binom{\mr{{\bf D}}}{\mr{{\bf H}}} = \Gamma^{-1}\chi \Gamma\binom{-\mr{{\bf E} }}{\mr{{\bf B}}} =  \begin{pmatrix}-\Sigma_1&0\\0&\Sigma_2 \end{pmatrix}
\binom{-\mr{{\bf E} }}{\mr{{\bf B}}}.
\enge\noi Defining the vectors $\d{{\bf D}} = \Sigma_1^{-1}\mr{{\bf D}},  \d{${\bf H}$} = \Sigma_2^{-1}\mr{{\bf H}}$,
it follows that
\bege
\binom{\d{{\bf D}}}{\d{{\bf H}}} = {{\begin{pmatrix}\Sigma_1^{-1}&0\\0&\Sigma_2\end{pmatrix}}} \binom{-\mr{{\bf D}}}{\mr{{\bf H}}} =
{{\begin{pmatrix}-1&0\\0&1\end{pmatrix}}} \binom{-\mr{{\bf E} }}{\mr{{\bf B}}}.
\enge\noi This implies that
\bege
\d{G} = \chi_0 {\mathring{F}},
\enge\noi where $\chi_0$ is the vacuum constitutive relations. Calculating the inverse maps, we obtain for any medium the constitutive
relation
\bege
G = \chi^{} F
\enge \noi where 
\bege
\chi^{} = \Gamma\Lambda\chi_0\Gamma^{-1}\enge\noi The constitutive tensor associated to the 32 crystal classes presenting natural optical activity
is described uniquely from  $\chi_0$, i.e., from the spacetime metric, since
$$
\chi^{\lambda\nu\si\kappa}_0 = Y_0 \sqrt{g} (g^{\lambda\si} g^{\nu\kappa} - g^{\lambda\kappa} g^{\nu\sigma}) \cite{post,hehl2}
$$\noi Using coordinates we write
\bege
\boxed{G_{\mu\nu}=\frac{Y_0}{4}\sqrt{g}\; \ep_{\mu\nu\ap\be}(\Gamma^\dagger)^{\ap}_\delta\Gamma^\be_\rho\Lambda^\si_\theta(g^{\delta\theta}
g^{\rho\lambda} - g^{\delta\lambda}g^{\rho\theta})\; F_{\si\lambda}}
\enge\noi Note that the expression above is the constitutive relation for any crystalline material, and it depends only of
the matrix $\g$ (given at the end of Sec. 3, for all crystal classes), that describes optical natural activity.
Then it can be seen as the deformation of the metric of Minkowski spacetime into a metric of curved riemannian spacetime, since
in order to describe the constitutive relations of any crystalline medium we only need the metric of Minkowski spacetime.
 
In particular, it is also possible to express, from the Lorentzian metric of Minkowski spacetime, the constitutive tensor associated to the dielectric and magnetic Faraday
 rotations, and the natural optical activity in arbitrary rotational symmetric media. It is respectively given by the following  matrices:
\bege
{\fn{\begin{pmatrix}
-\vcx_{11}&0&0&0&0&0\\
0&-\vcx&-i\vcx_{23}&0&0&0\\
0&i\vcx_{23}&-\vcx&0&0&0\\
0&0&0&1/\mu&0&0\\
0&0&0&0&1/\mu&0\\
0&0&0&0&0&1/\mu\end{pmatrix}}},\quad
{\fn{\begin{pmatrix}
-\vcx&0&0&0&0&0\\
0&-\vcx&0&0&0&0\\
0&0&-\vcx&0&0&0\\
0&0&0&\zeta_{11}&0&0\\
0&0&0&0&\mu^{-1}&i\zeta_{23}\\
0&0&0&0&-i\zeta_{23}&\mu^{-1}\end{pmatrix}}},\enge\noi
\bege
{\fn{\begin{pmatrix}
-\vcx&0&0&i\g_{11}&0&0\\
0&-\vcx&0&0&i\g_{11}&0\\
0&0&-\vcx&0&0&i\g_{11}\\
i\g_{11}&0&0&1/\mu&0&0\\
0&-i\g_{11}&0&0&1/\mu&0\\
0&0&-i\g_{11}&0&0&1/\mu\end{pmatrix}}},\enge\noi where $\zeta_{ij} = (\mu^{-1})_{ij}$. Post proves [Po97] that
electromagnetic waves propagate with phase velocity $u$
given by
\bege
u = \pm((\vcx \pm \vcx_{23})\mu)^{-1/2},
\enge\noi (dielectric Faraday rotation),
\bege
u = \pm \sqrt{\frac{\zeta \pm \zeta_{23}}{\vcx}},
\enge\noi (magnetic Faraday rotation) and
\bege
u = \pm \frac{\g_{11}}{\vcx}\pm\sqrt{\frac{1}{\vcx\mu} + \frac{\g_{11}^2}{\vcx^2}},
\enge\noi (natural optical activity).

In the whole process described in this subsection, we only have accomplished conformal transformations in $\RR^{1,3}$.

\section*{Concluding Remarks}
We investigated the relation between electrodynamics in anisotropic material media and its analogous 
formulation in an spacetime, with non-null Riemann curvature tensor.  The propagation of 
electromagnetic waves in material media is proved to be analogous to consider the electromagnetic wave propagation in the vacuum, now 
in a curved spacetime, which is obtained by a deformation of the Lorenztian metric of Minkowski spacetime. Such process of performing 
deformations of the metric of Minkowski spacetime can be rigorously described using extensors. 
Also, there exists a close relation between Maxwell equations in curved spacetime
and in an anisotropic material medium, indicating that electromagnetism and spacetime properties are deeply related.
 Besides, the geometrical aspects of wave propagation can be described by an effective geometry which represents a modification of the Lorentzian metric 
of Minkowski spacetime.

We discussed the optical activity of a given material medium, closely related to 
topological spin, and the Faraday rotation, associated to topological torsion. Both quantities are defined in terms 
of the magnetic potential and the electric and magnetic fields and excitations.The existence of form fields that are closed,
 but {\it not} exact, gives rise to the monopole and solitons in fluids, concerning 
topological defects and turbulent non-equilibrium thermodynamics, exhaustively investigated 
by Kiehn \cite{kiehn}. In a forthcoming paper, since the integral over $\RR^3$ of the topological torsion spatial component, introduced by eq.(\ref{tt}),
is the writhe of a framed oriented link, it is possible to investigate link invariants in gauge theory, from the knot theory viewpoint.


\begin{thebibliography}{99}
\footnotesize

\bibitem{ton} Tonti E, \emph{On the geometrical structure of electromagnetism}, in  Ferraese G (editor), \emph{Gravitation, 
Electromagnetism and Geometrical Structures}, 281-308, Pitagora, Bologna 1996.

\bibitem{heav} Heaviside O, \emph{Electromagnetic Theory}, Ernest Benn, London 1925.


\bibitem{war1} Warnick K F, Selfridge R H and Arnold D V {\it
Teaching electromagnetic field theory using differential forms},
IEEE Trans. Educ. {\bf 40}, 53-68 (1997).

\bibitem{war2} Warnick K F, {\em A Differential
Forms Approach to Electromagnetics in Anisotropic Media}, PhD. Thesis, Department of
Electric Engineering and Computation, Brigham Young University 1997.

\bibitem{war3} Warnick K F and Arnold D V,
{\it Green forms for anisotropic, inhomogeneous media}, J.
Electromagnet Wave {\bf 11}, 1145-1164
(1997).

\bibitem{bal}  Baldomir D and Hammond P, \emph{Geometry of Electromagnetic Systems},  Clarendon Press, Oxford 1996. 

\bibitem{fre} Freire I L, {\it Aplications of Differential Forms to the Anisotropic Media Electrodynamics} (in Portuguese), 
M.Sc. Thesis, Department of Applied Mathematics, Unicamp, Campinas 2004.

\bibitem{hej}Helszajn J, {\it Principles of Microwave Ferrite Engineering}, Wiley-Interscience, London 1969.

\bibitem{punt} Puntigam R A, L\"ammerzahl C and Hehl F W, \emph{Maxwell's theory on a post-Riemannian spacetime and the equivalence principle},
Class. Quantum Grav. {\bf 14} 1347-1356 (1997).

\bibitem{scho} Sch\"onberg M, {\it Electromagnetism and gravitation}, Revista Brasileira de F\'{\i}sica, 91-122 (1971).

\bibitem{slg} Salingaros N, \emph{Electromagnetism and the holomorphic properties of spacetime}, J. Math. Phys. {\bf 22} (9) 1919-1925 (1981).

\bibitem{lor} de Lorenci V A and Souza M A, \emph{Electromagnetic wave propagation inside a material medium: an effective geometry interpretation},
Phys. Lett. B {\bf 512} 417-422 (2001).

\bibitem{bay}  Baylis W, {\it Electrodynamics: A Modern Approach}, Birkh\"auser, Boston (1999).


\bibitem{born}  Born M and  Wolf E, \emph{Principles of Optics: Electromagnetic Theory of Propagation, Interference and Diffraction of Light}, 
Cambridge Unive. Press, Cambridge 1999. 



\bibitem{jad} Jadczyk A Z, \emph{Electromagnetic Permeability of the Vacuum and Light-Cone Structure}, Bulletin de l'Academie Polonaise des Sciences,
Serie des Sciences Physiques et Astron. {\bf 18}, 91-94 (1979).

\bibitem{schou} Schouten J A, \emph{Tensor Analysis for Physicists}, Dover, New York 1989.


\bibitem{Gronw} Gronwald F, Muench U, Mac\'{\i}as A and Hehl F, {\it Volume elements of spacetime and a quartet of scalar fields},
[{\tt gr-qc/9712063}].

\bibitem{lounesto}  Lindell I and Lounesto P, \emph{Differentiaalimuodot S\"ahk\"omagnetiikassa}, Helsinki University of Technology,
 Electromagnetics Laboratory report, Espoo 1995.

\bibitem{misner} Misner C, Thorne K and Wheeler J, \emph{Gravitation}, Freeman, San Francisco 1973.

\bibitem{gross} Gross A and Rubilar G, {\it On the derivation of the spacetime metric from linear electrodynamics}, [{\tt gr-qc/0103016}].

\bibitem{max} Maxwell J C, {\it A Treatise on Electricity and Magnetism}, vols.1 and 2, Dover, New York 1954.

\bibitem{hehl1} Hehl F and Obukhov Y, {\it Spacetime metric from linear electrodynamics I, II} [{\tt gr-qc/9904067}], [{\tt gr-qc/9911096}].

\bibitem{hehl2} Hehl F and Obukhov Y, {\it A gentle introduction to the foundations of classical electrodynamics:
the meaning of the excitations and the field strengths}, [{\tt physics/0005084}].

\bibitem{hehl3} Hehl F and Obukhov Y, {\it On the energy-momentum current of the electromagnetic field in a pre-metric axiomatic
approach: I} [{\tt gr-qc/0103020}].

\bibitem{hehl4} Hehl F, Obukhov Y and Rubilar G, {\it Light propagation in generally covariant electrodynamics and the
Fresnel equation}, [{\tt gr-qc/0203096}].

\bibitem{hehl5} Hehl F, Obukhov Y and Rubilar G, {\it On a possible new type of a T odd skewon field linked to electromagnetism},
 [{\tt gr-qc/0203105}].

\bibitem{janc1} Jancewicz B,
 {\it A variable metric electrodynamics. The Coulomb and Biot-Savart laws in anisotropic media}, Annals of Physics, {\bf 245}, 227 (1996).

\bibitem{janc2} Jancewicz B, \emph{The extended Grassmann algebra in $\RR^3$}, chap. 28 in Baylis W E (editor),
\emph{Clifford (Geometric) Algebras with applications in Physics, Mathematics and Engineering}, Birkh\"auser, Berlin 1996.


\bibitem{hill1} Hillion P, {\it Manifestly covariant formalism for electromagnetism in chiral media}, Phys. Rev, {\bf R47}(2), 1365-1374 (1993).

\bibitem{hill} Hillion P, {\it Electromagnetism in anisotropic chiral media}, Phys. Rev, {\bf E47}(4), 2868-2873 (1993).


\bibitem{kiehn} Kiehn R M, (a) {\it Chirality and helicity vs. spin and torsion or differential topology and
electromagnetism}; (b) \emph{Topological Torsion, Pfaff Dimension and Coherent structures};
 (c) \emph{ Non-equilibrium and irreversible electromagnetism from a topological perspective}; 
 (d) \emph{A topological perspective of non-equilibrium electromagnetism}; 
 (e) \emph{The Photon Spin and other Topological Features of Classical Electromagnetism}; (f)  
\emph{Topological Defects, Coherent Structures and Turbulence}; (g) \emph{
Spinors, Minimal Surfaces, Torsion, Helicity, Chirality, Spin, Twistors, Orientation, Continuity, Fractals, Point Particles, Polarization, the Light Cone and the Hopf Map};
(h) \emph{Electromagnetic Waves in the Vacuum with Torsion and Spin}; (i) \emph{ 
Topological Torsion and Spin form Coherent Structures in Plasmas and electromagnetic media}; (j) 
\emph{Chirality and Helicity vs Topological Spin and Topological Torsion}; (k) 
\emph{Optical Vortices and Topological Torsion}; (l) {\it The chiral vacuum}, http://www.cartan.pair.com.

\bibitem{whi} Whitaker E T, \emph{A History of the Theories of Aether}, Dublin Univ. Press, Dublin (1910).

\bibitem{post} Post E J, {\it The constitutive map and some of its ramifications}, Ann. Phys, {\bf 71}, 497-518 (1972).


\bibitem{postl} Post E J, {\it Formal Structure of Electromagnetics}, Dover, New York 1997.


\bibitem{sal} Salingaros N, {\it Electromagnetism and the holomorphic properties of spacetime}, J. Math. Phys., {\bf 22}(9) 1919-1925
(1981).


\bibitem{bt}  Benn I and Tucker R, {\it An Introduction to Spinors and Geometry with applications in Physics}, Adam Hilger, Bristol 1987.


\bibitem{bur} Burke W, \emph{Applied Differential Geometry}, Cambridge Univ. Press, Cambridge 1985.

\bibitem{voi} Voigt W, {\it Lehrbuch der Kristallphysik}, B. Teubner, Leipzig 1910.

\bibitem{bate} Bateman H, \emph{The transformation of the electrodynamical equations}, Proc. London Math. Soc. {\bf 8}, 223-264 (1910).

\bibitem{weyl} Weyl H, {\it Symmetry}, Princeton University Press, Princeton, New Jersey 1989.

\bibitem{tei} Teixeira F L and Chew W C, \emph{Unified analysis of perfectly matched layers using differential forms},
 Microwave and Optical Technology Letters {\bf 20} (2), 124-126 (1999).

\bibitem{rota} Barnabei M, Brini A and  Rota G-C, {\it On the exterior calculus of invariant theory}, J. Algebra, {\bf 96}, 120-160 (1985).

\bibitem{bf1} Conradt O, \emph{A treatise on quantum Clifford algebras}, Ph.D. thesis, Konztanz University, Konstanz 1999 [{\tt math-ph/0202059}].

\bibitem{bf2} Conradt O, \emph{Mechanics in Space and Counterspace}, J. Math. Phys. {\bf 41}, 6995-7028 (2000).







\end{thebibliography}
\end{document}